\documentclass{optica-article}

\journal{opticajournal} % for journals or Optica Open

\articletype{Research Article}

\usepackage{amsmath}
\usepackage{amsfonts}
\usepackage{bm}
\usepackage{lineno}
\begin{document}

\title{Nonlinear techniques for few-mode wavefront sensors}

\author{Jonathan Lin,\authormark{1}}
\author{Michael P. Fitzgerald,\authormark{1}}
\authormark{1} University of California, Los Angeles

\email{\authormark{*}jon880@astro.ucla.edu} %% email address is required; see note below about the corresponding author designation

% use {asbstract*} to suppress the copyright line. Copyright information will be added in production

\begin{abstract*} 
We present several nonlinear wavefront sensing techniques for few-mode sensors, all of which are empirically calibrated and agnostic to the choice of wavefront sensor. The first class of techniques involves a straightforward extension of the linear phase retrieval scheme to higher order; the resulting Taylor polynomial can then be solved using the method of successive approximations, though we discuss alternate methods such as homotopy continuation. In the second class of techniques, a model of the WFS intensity response is created using radial basis function interpolation. We consider both forward models, which map phase to intensity and can be solved with nonlinear least-squares methods such as the Levenberg-Marquardt algorithm, as well as backwards models which directly map intensity to phase and do not require a solver. We provide demonstrations for both types of techniques in simulation using a quad-cell sensor and a photonic lantern wavefront sensor as examples. Next, we demonstrate how the nonlinearity of an arbitrary sensor may studied using the method of numerical continuation, and apply this technique both to the quad-cell sensor and a photonic lantern sensor. Finally, we briefly consider the extension of nonlinear techniques to polychromatic sensors.
\end{abstract*}

%%%%%%%%%%%%%%%%%%%%%%%%%%  body  %%%%%%%%%%%%%%%%%%%%%%%%%%
\section{Introduction}

Linear wavefront sensing methods have long been the standard for astronomical adaptive optics (AO) due to their speed and simplicity. In exchange, such methods lose dynamic range as the intensity response of the wavefront (WFS) deviates from linearity, which can happen rapidly for sensors such as the unmodulated pyramid \cite{pyr,unmodpyr}, photonic lantern (PL;  \citenum{Leon-Saval:05,Lin:22,Lin:23}), and Zernike WFS \cite{zk}. As such, there has been a growing movement in the community to investigate and implement ``nonlinear'' sensing methods, an umbrella term applied to any phase retrieval method that uses more information than just the Jacobian and reference intensity image of the WFS. The need for such techniques has been underscored by the recent pushes for wavefront sensing in the final focal plane, in order to mitigate non-common-path aberrations (NCPAs; \citenum{Martinez:12:NCPA1}) and low-wind-effect aberrations \cite{NDiaye:18}, as well as the possibility of wavelength-dispersed WFSs, potentially enabled by microwave kinetic inductance detectors (MKIDs; \citenum{mkid}) and/or astrophotonic devices \cite{roadmap}. A final motivation for nonlinear sensing is that it will expedite alignment and NCPA compensation of AO systems, especially for WFSs with small dynamic range.
\\\\
Existing non-linear phase retrieval methods include phase diversity \cite{COFFEE,FF,DRWHO}, Gerchberg-Saxton algorithms \cite{Gerchberg72,kyohoon}, and most recently, neural networks \cite{Landman:20,Norris:20,tomeu}. Some types of WFSs, such as the pyramid and Zernike sensors, also have sensor-specific strategies (e.g. \citenum{Frazin:18,Hutterer:18,chamb} for the pyramid and \citenum{haffert} for the Zernike WFS). Broadly speaking, methods are often divided into ``model-based'' and ``model-free'' methods, though in the authors' opinion both classes of methods use models, with the main difference being whether the model is analytically developed or empirically calibrated; the classic linear sensing scheme is an example of the latter. The purpose of this article is to discuss alternative, empirically calibrated or ``data-driven'' retrieval methods, and to present numerical demonstrations. To offset the additional complexity of nonlinear sensing, we first apply our methods to few-moded ($\lesssim 10$ mode) WFSs: this includes sensors which physically cannot sense large numbers of modes (e.g. the PL) as well as sensors whose outputs are projected to a few-mode subspace (e.g. a pyramid sensor restricted to the first few control modes) and any sensors composed of many smaller few-mode sensors (e.g. Shack-Hartmann). We leave the extension of the concepts presented here to higher-mode-count sensors for future work.
\\\\
This article is divided into four subsequent parts. In \S\ref{sec:framework}, we establish the notation that will be used throughout the rest of the work. In \S\ref{sec:model} we discuss the different methods by which the nonlinear response of a WFS can be calibrated, including power-series expansion as well as interpolation using radial basis functions or polyharmonic splines. In \S\ref{sec:inv} we discuss how models can be solved, as well as the possibility of ``backwards'' models where the calibration procedure takes WFS intensities as inputs and phase aberrations as outputs. In \S\ref{sec:num} we present numerical demonstrations of these methods, and additionally show how the technique of numerical continuation can be used to better understand WFS nonlinearity. We also briefly discuss polychromatic wavefront sensing, leaving a more detailed analysis for future work.

\section{Framework}\label{sec:framework}
We represent a telescope + WFS system as a vector-valued nonlinear function which maps pupil-plane wavefront phases to intensity values. Extending the ``response matrix'' terminology from linear phase retrieval, we will refer to this nonlinear function as the ``response map'' of the WFS; the response map may also be loosely thought of as the WFS's ``transfer function'' in the sense of control theory. Denote the monochromatic response map for an arbitrary WFS as $\mathcal{A}_\lambda$, which takes both a real scalar field $\phi$ and a wavelength $\lambda$, and returns a vector corresponding to the $N$ intensity outputs of the WFS:
\begin{equation}\label{eq:1}
    \mathcal{A}_\lambda \left[\phi(x,y,\lambda)\right] = \bm{p}_\lambda. 
\end{equation}
Here, $\phi(x,y,\lambda)$ is the phase at wavelength $\lambda$ measured over the pupil plane. Note that $\lambda$ appears twice: once representing the chromaticity of the phase, and once representing the chromaticity of the response map. We can approximate a polychromatic WFS as a series of such equations for each $\lambda$. In the case where the phase chromaticity dominates, the polychromatic WFS acts as phase-diverse monochromatic sensor that observes the same phase aberration ``shape'' at different amplitudes. However, if the ``intrinsic'' WFS chromaticity dominates, the WFS may see orthogonal sets of phase modes at different wavelengths. For brevity, we will drop $\lambda$ going forwards, and only re-add it when discussing polychromatic sensors.
\\\\
The goal of wavefront sensing is to robustly solve the above equation on a timescale faster than that of the wavefront aberrations. Solutions are often approximated by expanding the aberration $\phi$ in a finite-dimensional and incomplete modal basis 
\begin{equation}
    \phi(x,y) = \sum_j a_j m_j(x,y)
\end{equation}
where the modes $m_j(x,y)$ are usually the ``most important'' modes (``control modes''), as determined by a singular value decomposition, and the $a_j$ are the mode amplitudes. The output vector $\bm{p}$ may also be projected to the subspace corresponding to the image of the selected phase modes. We denote this modal-basis, vector-valued function as $\mathcal{T}$, whose scalar components are
\begin{equation}
    \mathcal{T}_i \left[\bm{a}\right] = p_i.
\end{equation}
Assuming an incomplete modal basis, $\mathcal{T}$ is an approximation of $\mathcal{A}$. For $M$ modes and $N$ WFS outputs, $\mathcal{T}$ maps $\mathbb{R}^M$ onto some subset of $\mathbb{R}^N$ corresponding to the range of intensities that the WFS can produce. From here, we would ideally obtain a relation of the form
\begin{equation}
    a_j = \mathcal{T}_j^{-1}\left[ \bm{p}\right].
\end{equation}
However, because there is no guarantee that $\mathcal{T}$ is bijective, there is usually no {\it single-valued} inverse function $\mathcal{T}^{-1}$ which satisfies the above over the entire range of $\mathcal{T}$; the notation $\mathcal{T}^{-1}\left[\bm{p}\right]$ instead denotes the {\it preimage} of the intensity $\bm{p}$. A typical function which shows such behavior is the complex exponential, $e^{z}$ for $z\in \mathbb{C}$, which is single-valued but whose inverse is the multi-valued complex logarithm: ${\rm log}(z) =  \{ {\rm ln} \, |z| + i({\rm Arg} \, z + 2\pi n)$ | $n\in \mathbb{Z}$\}. Here, ${\rm Arg}(z)$ denotes the principal value complex argument of $z$, which is restricted to take values in the interval $(-\pi,\pi]$. In the context of interferometry, the failure for $e^z$ to admit a single-valued inverse is referred to as ``phase wrapping''.
\\\\
In \S\ref{ssec:char} we will explicitly investigate how $\mathcal{T}$ fails to admit a single-valued inverse. To do so, it will be useful to embed $\mathcal{T}$ into the function $\mathcal{F}:\mathbb{R}^{M+N}\rightarrow \mathbb{R}^{N}$ with
\begin{equation}\label{eq:embed}
    \mathcal{F}\left[\bm{a},\bm{p}\right] = \mathcal{T} \left[\bm{a}\right] - \bm{p}.
\end{equation}
Pairs of corresponding intensities and phases fall along the surfaces with $\mathcal{F}\left[\bm{a},\bm{p}\right] = \bm{0}.$

\section{WFS models} \label{sec:model}
In this section, we discuss some methods which can be used to empirically model the nonlinear WFS response represented by equation \ref{eq:1}: the first step in a sensor-agnostic nonlinear phase retrieval scheme.
\subsection{Power series representation}
\subsubsection{Linear approximation}
To begin, we review the standard linear approach to wavefront sensing. A Taylor expansion to first order about a reference aberration $\bm{r}$ yields
\begin{equation}
    p_i(\bm{r}+\bm{x}) - p_i(\bm{r}) = \partial_j p_i\big|_{\bm{r}}x_j
\end{equation}
where $\partial_i \equiv \partial /\partial x_i$ and the vertical bar denotes the point of evaluation. The derivative term is typically measured using centered finite difference; thus, an $M$-moded sensor requires 2$M$ probes to fully calibrate the linear model. Denoting the slope matrix (the Jacobian) as $T_{ij} \equiv \partial_j p_i\big|_{\bm{r}}$, we can write the above relation as 
\begin{equation}
    \bm{p}(\bm{r}+\bm{x}) -\bm{p}(\bm{r})  \approx T \bm{x}
\end{equation}
which is inverted as
\begin{equation}
    \bm{x} =T^+ \left[ \bm{p}(\bm{r}+\bm{x}) - \bm{p}(\bm{r})\right]
\end{equation}
where $T^+$ is the inverse or pseudo-inverse of $T$. The map $T$ is the linearization of $\mathcal{T}$.

\subsubsection{Higher-order Taylor expansion}
The natural extension of the linear technique is a power series expansion to higher order:
\begin{equation}\label{eq:tensor}
    p_i(\bm{r}+\bm{x}) = p_i(\bm{r}) + \partial_j p_i\big|_{\bm{r}}x_j + \dfrac{1}{2}\partial_j \partial_k p_i\big|_{\bm{r}} x_j x_k + \dfrac{1}{6} \partial_j \partial_k\partial_l p_i\big|_{\bm{r}} x_j x_k x_l +...
\end{equation}
Repeated indices within the same term imply summation. For a few-moded WFS, we can truncate the power series beyond linear order. For instance, calibration up to second order requires an additional $2M^2$ probes: off-diagonal terms in the Hessian each involve four phase probes corresponding to the vertices of a small square centered on the reference phase, while the diagonal terms each require three measurements, one of which is supplied by the reference response of the WFS. In principle, truncation can occur at arbitrary order, though eventually calibration will take an unfeasible amount of time. Also, depending on the convergence of the series, higher order terms may be detrimental to dynamic range.

\subsubsection{Higher-order expansion from complex response map}
While the response map $\mathcal{T}$ is nonlinear, we note that the response map relating the complex-valued electric field in the pupil plane to that at the WFS output is linear. If this map can be measured (e.g. with phase diversity, off-axis holography, or a spatial light modulator), then it can be used to determine all higher-order tensors of the Taylor expansion. We direct the reader to \cite{Lin:22} for more details.

\subsection{Radial basis function interpolation}
Instead of expanding the nonlinear WFS response in terms of polynomials, another approach to build a model of the nonlinear response, often used in unstructured data analysis, is to expand in terms of radial basis functions (RBFs; \citenum{RBF}). An RBF is a function $\theta:\mathbb{R}^M \rightarrow \mathbb{R}$ whose value depends only on the distance to some reference point. For example, the Gaussian RBF is 
\begin{equation}
    \theta(\bm{x},\bm{r}) = e^{-\epsilon^2||\bm{x}-\bm{r}||^2}
\end{equation}
where $\epsilon$ is called the ``shape parameter'' of the RBF, $\bm{r}$ is the reference point, and $||\bm{x}||$ is the norm of $\bm{x}$. A nonlinear function $f:\mathbb{R}^M\rightarrow \mathbb{R}$ can be expanded as 
\begin{equation}
    f(\bm{x}) = \sum_i w_i \theta(\bm{x},\bm{r}_i)
\end{equation}
where the $w_i$ are the weights and the $\bm{r}_i$ are the ``control points'' of the interpolation. Given a set of $P$ control points, one can determine the weights by solving the $P\times P$ linear system
\begin{equation}
    \begin{bmatrix}
        \theta(\bm{r}_1,\bm{r}_1) & \theta(\bm{r}_1,\bm{r}_2)  & \dots \\
        \theta(\bm{r}_2,\bm{r}_1) & \theta(\bm{r}_2,\bm{r}_2) & \dots \\
        \vdots & \vdots & \ddots
    \end{bmatrix}
    \begin{bmatrix}
        w_1 \\
        w_2 \\
        \vdots
    \end{bmatrix}
    = \begin{bmatrix}
        f(\bm{r}_1) \\
        f(\bm{r}_2)  \\
        \vdots
    \end{bmatrix}.
\end{equation}
To extend the technique to vector-valued nonlinear functions, the simplest approach (and the approach adopted later in this work) is to apply the scalar interpolation technique to each output dimension independently; more advanced extensions are discussed in \S\ref{ssec:interp2}.
Other RBFs besides the Gaussian are also used; notably, functions with compact support will yield sparse linear systems. A related technique is polyharmonic spline interpolation \cite{spline}, which augments the RBFs with a polynomial term. For instance, we may add a linear term as
\begin{equation}
    f(\bm{x}) = \sum_i w_i \theta(\bm{r}_i,\bm{x}) + \bm{v}^T \cdot \begin{bmatrix}
        1 \\
        \bm{x}
    \end{bmatrix}.
\end{equation}
Here, $\bm{v}$ is a column vector of linear polynomial weights, with dimension $M+1$. Like with pure RBF interpolation, the weights for the polyharmonic splines are determined by solving a linear system. The default interpolation scheme implemented by \texttt{scipy}'s \texttt{RBFinterpolator}, which we use for numerical tests in \S\ref{sec:num}, is a polyharmonic spline interpolation using the RBF $\theta(r) = r^2 \log r$ and a linear polynomial term. 
\\\\
The simplest way to select the control points for an RBF model is to randomly sample from a subset of the phase aberration space --- for example, a ball with maximum radius corresponding to the sensor's expected dynamic range. We would then experimentally measure the WFS response at these points and determine the interpolation weights as above. If first-order derivative information is also required, we can additionally develop an RBF model of the Jacobian. In \S\ref{sec:num}, we present some initial numerical results on how many control points are required to obtain a good fit. The number of control points can be reduced through greedy algorithms, in which new sample points are repeatedly inserted where the model error is highest \cite{greedu}.

\subsection{Other interpolation methods}\label{ssec:interp2}
There are many other options for data-driven interpolation models other than the RBF method mentioned above, which we do not consider but could be explored in future works. For instance, a common method used to interpolate univariate curves in computer graphics is basis spline (b-spline; \cite{UnserSpline}) interpolation. B-splines have finite support and can be extended to higher-dimensional spaces by forming multi-dimensional basis functions from the tensor products of the univariate b-splines. Another technique often used in statistics and machine learning is Gaussian process modeling \cite{SCHULZ20181}, a non-parametric Bayesian regression technique which considers the space of functions as an ``infinitely'' multivariate normal distribution whose covariance is determined by a kernel function. This distribution is then conditioned on observed data, and the mean of the conditioned distribution is taken as the interpolant. Similar to RBF interpolation, the standard formulation of Gaussian process modeling treats univariate data, but the technique can be extended to the multivariate case by applying univariate method to each output independently. Alternatively, more advanced ``multi-output'' Gaussian processes, which involve prescriptions to model the output covariances, are possible \cite{mogp}.
\\\\
Notably, there is a deep connection between RBF interpolation, spline interpolation, and Gaussian process modeling, all of which are a type of non-parametric regression. For instance, spline interpolation can be obtained as a specific case of Gaussian process modeling through the choice of kernel \cite{spl}; similarly, RBF interpolation and Gaussian process modeling are equivalent if the RBF is taken as the kernel of a Gaussian process, though this may not always be warranted (for instance, not all RBFs are valid covariances). All of these techniques are a form of functional regression in Reproducing Kernel Hilbert Spaces \cite{rhks}. We leave further investigations of alternate interpolation methods beyond RBF interpolation to future work.

\subsection{Neural network models}
Similar to a non-parametric interpolation method, neural networks provide a way to model an arbitrary nonlinear function. The calibration process is also similar: we must provide a training data set of input phase aberrations and output intensities. The weights of the neural network are then determined via gradient descent. The amount of time required for RBF interpolation models and neural network models seems comparable: \cite{Norris:20} implements a neural network reconstructor for a few-mode PL sensor which is trained using calibration datasets containing $10^3-10^4$ points, similar to what we use in our simulations in \S\ref{sec:num}. However, we note that as of now RBF interpolation is simpler to implement.  We direct the reader to \cite{Norris:20,Landman:20,tomeu} for more details. 

\section{Inversion strategies}\label{sec:inv}
Given a model of the WFS response in the neighborhood of some reference phase $\bm{r}$, we seek a way to find the preimages of the observed intensity patterns. We discuss options in the next sections.

\subsection{Standard methods}\label{ssec:stdmethod}
Briefly, we mention that there are a number of well-developed nonlinear equation solvers including conjugate gradient methods (e.g. Newton-CG), nonlinear least-squares methods such as Levenberg-Marquardt \cite{Leve44}, and more general trust region methods. For few modes, these methods may run quickly enough for real-time AO.

\subsection{Solution to the Taylor model --- successive approximations}
In the case of Taylor expansion, there is another method based on successive approximations \cite{Kantorovitch1939}. We seek to solve the following equation for $\bm{x}$ given a vector $\bm{\rho}$ and tensors $(B,C,D, ...)$:
\begin{equation} \label{eq:taylor}
    \rho_i(\bm{r},\bm{x}) = B_{ij}x_j + C_{ijk} x_j x_k + D_{ijkl} x_j x_k x_l + ...
\end{equation}
In the context of WFSs, $\rho_i$ is the reference-subtracted intensity response. To first order, we have 
\begin{equation}
    \bm{x}^{(1)} = B^+\bm{\rho} + o(\bm{x}^2).
\end{equation}
The second order equation is approximated by a partial substitution:
\begin{equation}
    \rho_i \approx \left[ B_{ij}  + C_{ijk} x_k^{(1)}\right] x_j = \left\{ B_{ij}  + C_{ijk}\left[ B^{+}_{km} \rho_m+o(\bm{x}^2)\right]\right\} x_j \equiv B^{(2)}_{ij}\left[ x_j + o(\bm{x}^3)\right]
\end{equation}
and so the second-order approximant to $\bm{x}$ is 
\begin{equation}
    \bm{x}^{(2)} = B^{(2)+}\bm{\rho} + o(\bm{x}^3),
\end{equation}
and so on. Note that if $\bm{x}^{(1)}$ is 0, all $x^{(n)}$ are 0: we cannot sense modes in the nullspace of $B$. 
\subsection{Radial basis function models}
There are several possibilities in solving interpolated models such as those constructed from RBFs. For one, the model of the WFS response map $\mathcal{T}$ can be combined with one of the nonlinear solving methods mentioned in \S\ref{ssec:stdmethod}, e.g. by minimizing the scalar quantity
\begin{equation}
    ||\mathcal{T}[\bm{a}] - \bm{p}||
\end{equation}
over aberration amplitude $\bm{a}$, given WFS image $\bm{p}$.
Models of the Jacobian of $\mathcal{T}$ may also enable gradient descent methods as well as phase retrieval models based on local linearization of the WFS \cite{jacrecon}, and can be used at the very least to speed up nonlinear solvers. However, the simplest scheme, akin to what is often done with neural network WFS models, is to construct the model ``in reverse'', swapping the inputs and outputs during the calibration process so that the RBF model returns phase aberrations as a ``function'' of WFS intensities. As mentioned in \S\ref{sec:framework} this will not work over arbitrarily large regions of phase aberration space, so control points must be placed in smart manner. We discuss this more in \S\ref{ssec:char}, but leave the development of an optimal method for WFS control point placement to future work. 

\subsection{Numerical continuation}
The class of numerical continuation methods \cite{Allgower1990} aim to solve a system of nonlinear equations by starting with a different system admitting known solutions, and then continuously transforming the starting system into the one in question. As the system is transformed, the solutions, which will also vary continuously, are tracked, so that when the transformation is complete we are left with the solutions to the target system. One example is homotopy continuation for polynomial root-finding: suppose we want the roots of a polynomial $P(\bm{x})$. We begin with a polynomial $P_0(\bm{x})$ whose roots are known; preferably, the number of roots in the starting polynomial should also match the number of expected roots in the target polynomial, otherwise we must augment our numerical continuation methods with {\it bifurcation theory} \cite{bifurc}, which is outside the scope of this paper. We then embed both systems in a homotopy class, for instance
\begin{equation}
    H(\bm{x},s) = (1-s) P_0(\bm{x}) + s P(\bm{x}),
\end{equation}
where $s \in [0,1]$. Note that the choice of homotopy is not unique, and can strongly impact the quality and speed of the solving process. Finally, the solutions $\bm{x}_0$ for which $H(\bm{x}_0,s)=0$ are tracked as $s$ is increased to 1. Typically, this tracking is done with ``predictor-corrector'' methods such as pseudo-arclength continuation \cite{pseudoarc}, which is briefly introduced in Appedix \ref{ap:pseudo}.
\\\\
Continuation methods for polynomials are well-developed with multiple existing software packages, making homotopy continuation a plausible method to solve the Taylor polynomial equation, at least in the case of few-mode sensors. Numerical continuation is also applicable for non-polynomial systems. Consider the embedded WFS response map in equation \ref{eq:embed}. Any curve in ``intensity space'', which we define as the image of the phase aberration space with respect to $\mathcal{T}$, implicitly defines a homotopy class of functions. By constructing a curve from some reference intensity $\bm{p}_0$ with known reference phase $\bm{r}$ to the observed intensity $\bm{p}$ and applying numerical continuation, we may determine the phase aberration solution $\bm{r}+\bm{x}$. If $\bm{p}_0$ is associated with multiple phase aberrations $\bm{r}_i$, then numerical continuation from each $\bm{r}_i$ will yield a different phase solution for $\bm{p}$.

\section{Numerical results}\label{sec:num}
\subsection{Method}
We simulate a 10-m telescope with an unobstructed circular aperture using the Python package \texttt{hcipy} \cite{hcipy}. We fix the wavelength to 1.55 $\mu$m and set a focal ratio of 4. As an initial demonstration, we then couple the optical system to a quad-cell tip-tilt sensor composed of four square pixels each with side length 10 $\mu$m. The four outputs of the quad cell are combined into two values in the usual bi-cell configuration, i.e. by taking the difference in flux between the left and right halves of the sensor, and the difference between the top and bottom halves, and then normalizing both differences by the summed flux of all four pixels. We chose a quad-cell as a first test because it is simple and demonstrates known non-linearity, and because techniques applied to quad-cells may be extended to Shack-Hartmann sensors. Later, we simulate a few-mode sensor based on a 6-port PL, which was modeled using the package \texttt{cbeam} \cite{cbeam}. The phase aberration space of this sensor was restricted to the first 5 non-piston Zernike modes (Noll indices 2-6, corresponding to tilt, defocus, and astigmatism). The lantern has an output single-mode core size of 2.2 $\mu$m, and the geometry of the input is similar to that of a step-index few-mode fiber with a core radius of 10 $\mu$m. The lantern tapers by a factor of 8 over a length of 4 cm.
\\\\
We tested two classes of phase retrieval methods. The first class is based on solving the Taylor polynomial equation using successive approximations; for reference, calibration up to third order for the 5-mode PL sensor required 285 deformable mirror probes. We also tested methods based on RBF interpolation. The first variant combines a forward model of the WFS response constructed from RBFs (and a linear term) with a nonlinear least-squares solver; we call this method ``RBF-LS'' for ``RBF least-squares''. In the second, we evaluate an inverse model of the WFS response, constructed from the same type of RBFs; we call this method ``RBF-I'' for ``RBF inverse''. In the last method, we solve the RBF-LS model using numerical continuation.
\\\\
Phase retrieval schemes were implemented in Python using \texttt{numpy} and \texttt{scipy}. In particular, for RBF interpolation and nonlinear solving, respectively, we used the functions \texttt{RBFinterpolator} and \texttt{least\_squares} as implemented by \texttt{scipy v1.14.0}. For the quad-cell, both the RBF-LS and RBF-I methods used the same set of 100 control points in phase space, which were randomly sampled in a ball of radius 2 radians RMS. Samples were selected uniformly in radius, so that there are more points near the origin. For the lantern sensor, we used 2000 control points sampled in a ball of radius 1 radians RMS. We later found that it may be possible to reduce the number of control points significantly while maintaining reasonable accuracy; see \S\ref{ssec:numpoints}. A simple code was written for numerical continuation, based off the method of pseudo-arclength continuation (Appendix \ref{ap:pseudo}).

\subsection{Phase retrieval comparisons}\label{ssec:phaseretrieval}
\begin{figure}
    \centering
    \includegraphics[width=\textwidth]{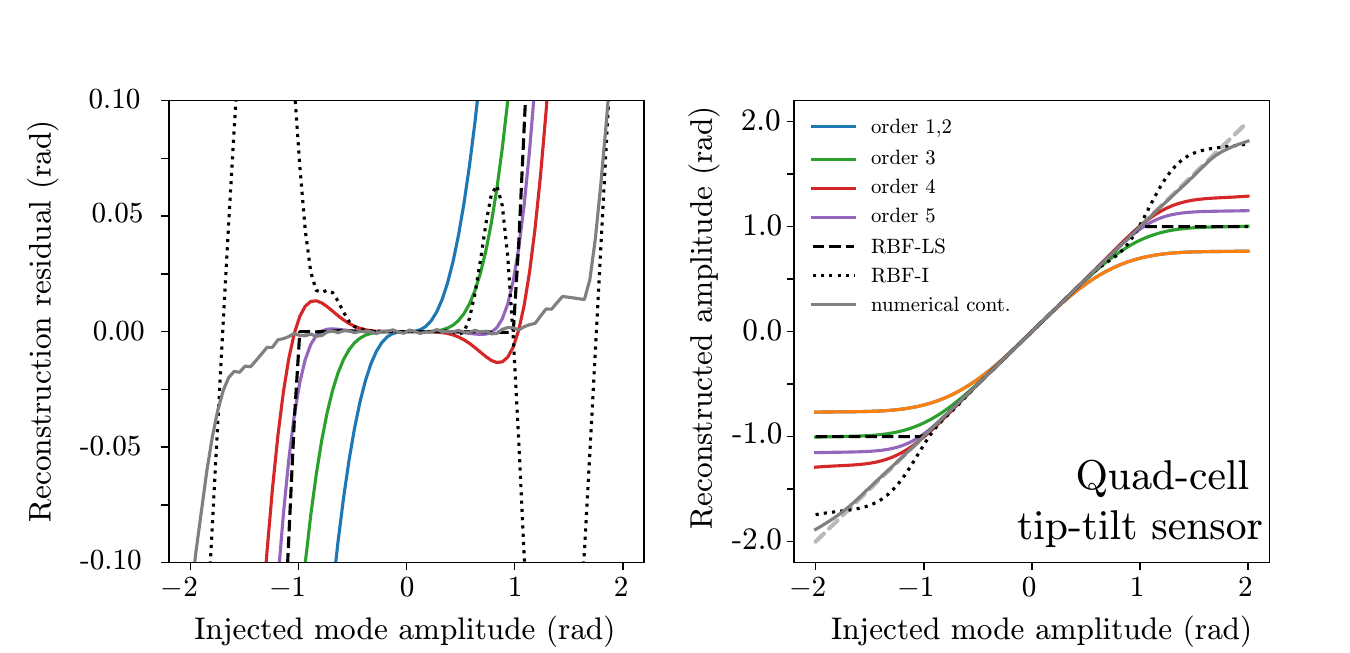}
    \caption{Left: residual of different phase retrieval schemes as a function of injected tilt for our fiducial quad-cell sensor. Order 1, 2, and so on correspond to Taylor expansion phase retrieval methods of the specified order. The ``RBF-LS" curve corresponds to solving a forward model of the WFS built using RBF interpolation, while the ``RBF-I'' curve corresponds to direct evaluation of a backward model of the WFS, also built with RBFs. Right: reconstructed tilt amplitude as a function of injected amplitude. The dashed grey line $y=x$ represents the behavior of a perfect of sensor.}
    \label{fig:quad}
\end{figure}

\begin{figure}
    \centering
    \includegraphics[width=\textwidth]{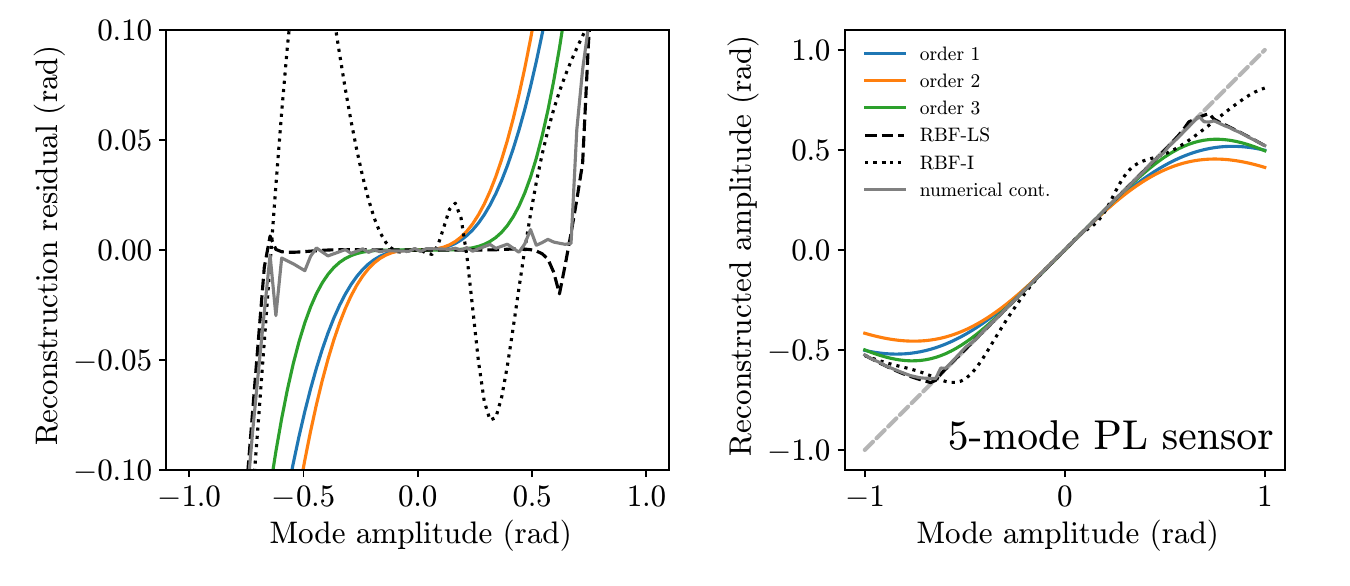}
    \caption{Left: residual of different phase retrieval schemes as a function of injected aberration mode amplitude, for our fiducual 5-mode PL sensor. In this particular plot, we chose to vary only the amplitude in Zernike mode 2 (tilt). Orders 1, 2, and 3 correspond to linear, quadratic, and cubic phase retrieval using Taylor approximation methods. The ``RBF-LS" curve corresponds to solving a forward model of the WFS built using RBF interpolation, while the ``RBF-I'' curve corresponds to direct evaluation of a backward model of the WFS, also built with RBFs. Right: reconstructed tilt amplitude as a function of injected amplitude.}
    \label{fig:tilt}
\end{figure}

\begin{figure}
    \centering
    \includegraphics[width=\textwidth]{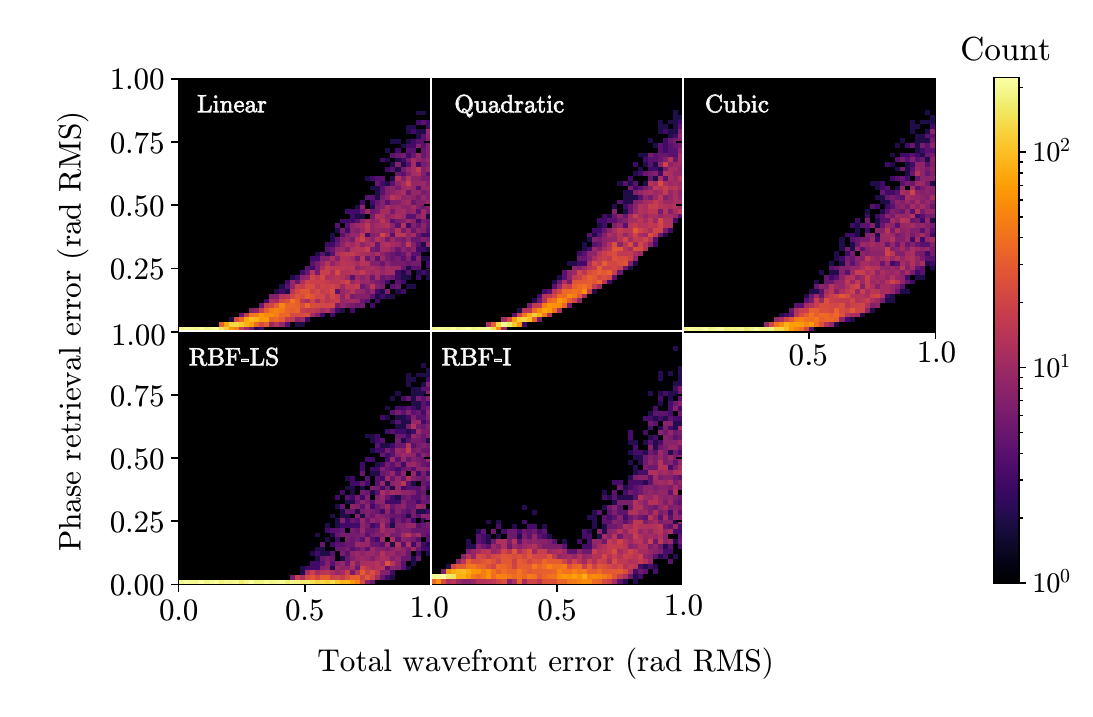}
    \caption{2D histograms plotting total phase retrieval error on the vertical axis and input RMS WFE on the horizontal axis, for different retrieval methods combined with the 5-mode PL sensor. Each plot was produced using 10000 randomly sampled phase aberrations within a ball of radius 1 radian RMS.}
    \label{fig:heatmap}
\end{figure}

In this section we present performance comparisons for the different phase retrieval methods mentioned in \S\ref{sec:model} and \S\ref{sec:inv}. We first injected varying amounts of a single aberration into the quad-cell system and attempted phase retrieval. Figure \ref{fig:quad} (left) shows the residual of the phase retrieval estimate for each method, as a function of injected tilt, while Figure \ref{fig:quad} (right) plots the retrieved tilt amplitude as a function of the injected tilt; the dashed grey line $y=x$ represents a perfect WFS. Comparing the Taylor expansion methods, we note that the quadratic method offers no benefit, and that the largest gains in accuracy come from odd orders; this is expected because the highest order term of an even Taylor expansion is symmetric about the expansion point, and is likely not as useful for sensing. The best performers were the numerical continuation and RBF-I methods. Defining the dynamic range as the region where the absolute value of the phase retrieval error is less than 0.1 rad, the numerical continuation method more than doubles the dynamic range of the linear method. However, numerical continuation is quite slow, which we discuss at the end of this section.
Next, we applied the phase retrieval methods to the 5-mode PL sensor. Figure \ref{fig:tilt} (left) shows the error in the phase retrieval estimate for each method, specifically as a function of injected tilt, while Figure \ref{fig:tilt} (right) plots the retrieved tilt amplitude as a function of injected tilt. Similar to the quad-cell, the largest gains in dynamic range from the Taylor methods come from the cubic model. The best performers were the RBF-LS and the numerical continuation methods. Interestingly, the RBF-I method performed worse than even the linear reconstructor for small amounts of aberration, but outperformed it for larger amounts. The increased error at small aberrations is likely due to our calibration data including pairs of degenerate or nearly degenerate phase aberrations, which make the model inconsistent; we discuss this more in \S\ref{ssec:char}.
\\\\
We next sampled 10,000 random phase aberrations with at most 1 radian of total RMS wavefront error (WFE), and compared retrieval errors at each sample point for the 5-mode PL sensor. Our results are shown in Figure \ref{fig:heatmap}, which plots a 2D histogram for each retrieval method with total input WFE on the horizontal axis and retrieval error on the vertical axis. The numerical continuation method is omitted because in this case it performs similarly to RBF-LS, but takes longer to run. The histograms show that the RBF-LS triples the dynamic range of the linear method, and that the RBF-I method is outperformed by all methods until around 0.5 radians of total RMS WFE. We also recorded each method's solving times for all 10,000 sampled aberrations. The linear, quadratic, and cubic methods took a total of 0.85, 1.85, and 3.12 seconds for all 10,000 solves. The RBF-LS and RBF-I methods took 39.49 and 0.59 seconds, respectively. We suppose that the RBF-I method only outperforms the linear method due to some additional overhead incurred by our Python code. Nevertheless, a better-calibrated RBF-I method with a smarter allocation of control points could enable extremely high-speed wavefront sensing at dynamic ranges much larger than the linear range of the sensor. Even the RBF-LS method could be used as-is for live correction of quasistatic NCPAs. For reference, our simple numerical continuation code takes $\sim 0.1$ s for a single solve. However, the continuation code was written in pure Python and is completely unoptimized. 

\subsection{Number of control points}\label{ssec:numpoints}
\begin{figure}
    \centering
    \includegraphics[width=\textwidth]{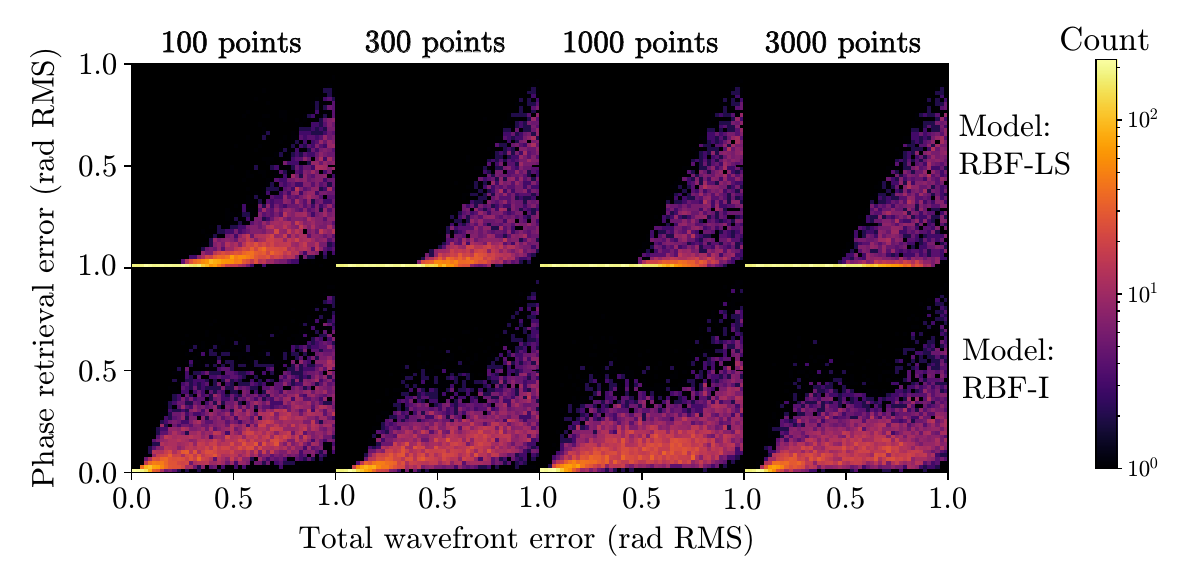}
    \caption{Comparison of RBF-based phase retrieval methods for different numbers of control points. We note the RBF-LS model begins seeing diminishing returns beyond 1000 control points, and still outperforms the linear model using only 100 control points. The RBF-I model appears almost insensitive to the number of control points.}
    \label{fig:numcomp}
\end{figure}

We next consider how many control points are required to obtain a good fit of WFS response using RBFs; this number directly correlates to the amount of time needed for calibration and determines how far the technique can be scaled in terms of the number of sensed modes. We repeat the random sample process used to generate Figure \ref{fig:heatmap} for the 5-mode PL sensor, this time attempting phase retrieval using models calibrated with different amounts of control points. Our results are shown in Figure \ref{fig:numcomp}. The phase retrieval error scales relatively weakly with the number of control points: even at only a 100 points, the RBF-LS method clearly outperforms the linear method, and beyond 1000 points there is almost no gain in accuracy. The RBF-I method shows an even weaker scaling. For the quad-cell, we find that this same threshold happens at around 100 control points. These two results suggest that the number of control points required for an RBF model scales sub-exponentially with dimensionality/mode count. 

\subsection{Characterization through numerical continuation}\label{ssec:char}
In this section we demonstrate that numerical continuation can be used to better understand WFS nonlinearity. The idea is as follows: we embed the nonlinear response map $\mathcal{T}$ in the function $\mathcal{F}$, as in equation \ref{eq:embed}. This function is $\mathbb{R}^N$-valued, with a domain that is some subset of $\mathbb{R}^N \times \mathbb{R}^M$ corresponding to the direct product of the space of WFS intensities and the space of phase aberrations. Then, we set $\mathcal{F}=0$, defining an implicit relation between the two spaces. Finally, we probe the intensity space along different paths, while computing the corresponding solution curves in phase space which together satisfy $\mathcal{F}=0$. Note that we may also do the reverse, and probe phase space along different paths while computing the corresponding WFS intensity. This is often done in WFS characterization and does not require numerical continuation. However, if the goal is to search for features such as phase degeneracies, or to otherwise understand how the intensity-to-phase map fails to be a single-valued function, this simpler method is less informative: for instance, we do not know {\it a priori} how phase space should be traversed to find degeneracies. In comparison, degeneracies are naturally identified using numerical continuation, as we will later see. 
\\\\
Because the curves have the potential to be quite complicated, we first use our method to display the nonlinearity of the simple two-mode quad-cell sensor from \S\ref{ssec:phaseretrieval}. Since this sensor maps aberrations in a 2D phase space to a 2D intensity space, we can simultaneously visualize paths in intensity space and their corresponding solution curves in phase space with side-by-side plots. This method can also be extended to higher dimensions by projecting onto different 2D or 3D subspaces; see \S\ref{ap:B} for an example with the 5-mode PL sensor.
\\\\
To begin our study, we start with a reference pair of points in intensity space and phase space corresponding to the reference phase and intensity; in our particular example the reference in each is [0,0], because a flat phase yields a zero measurement for both bi-cell summations. We then attempt to move in intensity space along a ray emanating from the reference intensity, and simultaneously use pseudo-arclength continuation to compute the corresponding phase aberration solutions. By repeating this process for a number of rays, we obtain the plots in Figure \ref{fig:nonlin}. Each line segment in the left panel represents a path in intensity space which was constrained to move along a ray, and defines a corresponding curve of solutions in phase space through the implicit relation $\mathcal{F}=0$.  Black circles in either space mark the location of simple folds, where the intensity response ``turns over'' as a function of arclength (i.e. goes from increasing to decreasing or vice versa). Such locations mark where the Jacobian loses rank. In phase aberration space, each aberration near a simple fold has a ``twin'' aberration on the other side of the fold which is mapped to the same intensity. Note that we only plot the first simple fold found during each numerical continuation to reduce clutter. From this plot, we make the following conclusions:
\begin{enumerate}
    \item Simple folds appear at around $2.5$ radians RMS of tilt. Beyond this threshold, the mapping from intensity space to phase space is no longer bijective. The simple folds divide phase space into two branches. The region bounded by simple folds and containing the origin is the principal branch, which roughly corresponds to the dynamic range of the sensor under closed-loop control and linear phase retrieval. %both branches are mapped to the same intensity space.
    \item Looking at the density of solution curves in phase space, we note that when increasing the total WFE, the quad-cell is less sensitive to diagonal tilts.
    \item The reference phase $[0,0]$ is unique: there is no other phase aberration that yields the same WFS response as the reference. Such aberrations, if they existed, would manifest as curve crossings in the full $M$-dimensional phase space. 
\end{enumerate}
As a more complex example, we repeat the above analysis for a two-mode tip-tilt sensor which uses a 3-port PL; the 3-dimensional output intensity vector is reference-subtracted and projected onto the two control modes with the two highest singular values so that our WFS maps $\mathbb{R}^2$ onto some subset of $\mathbb{R}^2$, like the quad-cell. In our case, this projection is also equivalent to a normalization of the WFS image. Figure \ref{fig:nonlin} shows our results, this time plotting all simple folds. We make several conclusions:
\begin{enumerate}
    \item There is a triangular region in phase space, bounded by simple folds, where the straight lines in intensity space are mapped to (mostly) straight lines. Outside this region we may find additional phase aberrations which are degenerate to the ones in this region.
    \item Both plots have three-fold rotational symmetry, like the 3-port PL. 
    \item We can explicitly see the 6 phase aberration solutions that are degenerate to the reference phase $[0,0]$. These points are represented by curve crossings in phase space. 
    \item The simple folds divide the space of phase solutions into 7 branches, with the principal branch corresponding to the central triangular region. Each branch maps to a different but overlapping region of intensity space. As a corollary, an intensity may have a different number of associated phase aberrations, depending on its location in intensity space.
    \item Regions devoid of curves in phase space indicate where our WFS is insensitive; this may change with the choice of control modes.
\end{enumerate}
We envision that similar analyses may be done for other WFSs. Such a method provides a means of computing the bounded region within which it is possible to formulate a one-to-one mapping between intensity and phase. Additionally, for any point in intensity space, it only takes two numerical continuations to compute the set of aberrations which map to that point. Finally, the most powerful aspect of this technique is the characterization of the WFS response into branches. If a one-to-one map can be constructed for each branch of the WFS response (i.e. by constructing a model at the reference phase and every point degenerate to it), and there is a regularizing method of predicting which solution branch we are in at any given time (e.g. using auxiliary AO telemetry), then accurate phase retrieval outside the limits set by degeneracy is possible.

\begin{figure}
    \centering
    \includegraphics[width=\textwidth]{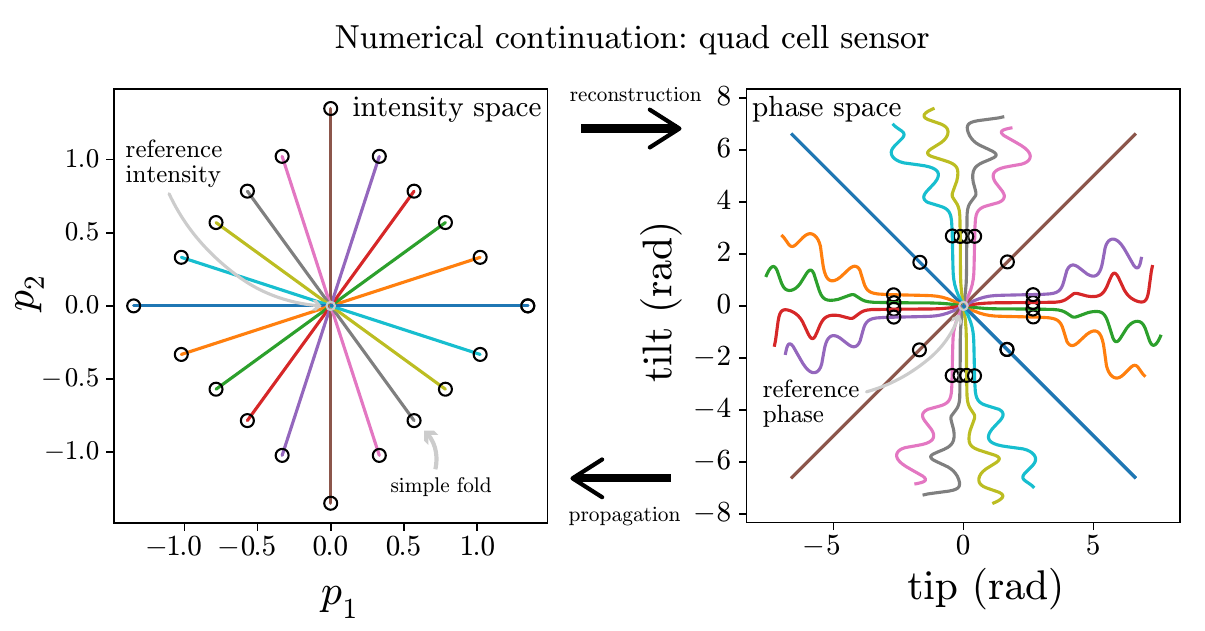}
    \caption{Numerical continuation plot for the quad-cell sensor from \S\ref{ssec:phaseretrieval}. Each colored line segment in intensity space (left) is mapped to a curve of the same color in phase space (right) by the nonlinear WFS response map. Black circles in either space mark the location of simple folds, where the intensity response ``turns over'' (i.e. goes from increasing to decreasing or vice versa). For clarity, only the first simple fold in each pair of curves is shown. Such locations indicate the presence of pairs of phase aberrations located on either side of the fold which are mapped to the same WFS intensity.}
    \label{fig:nonlinqc}
\end{figure}

\begin{figure}
    \centering
    \includegraphics[width=\textwidth]{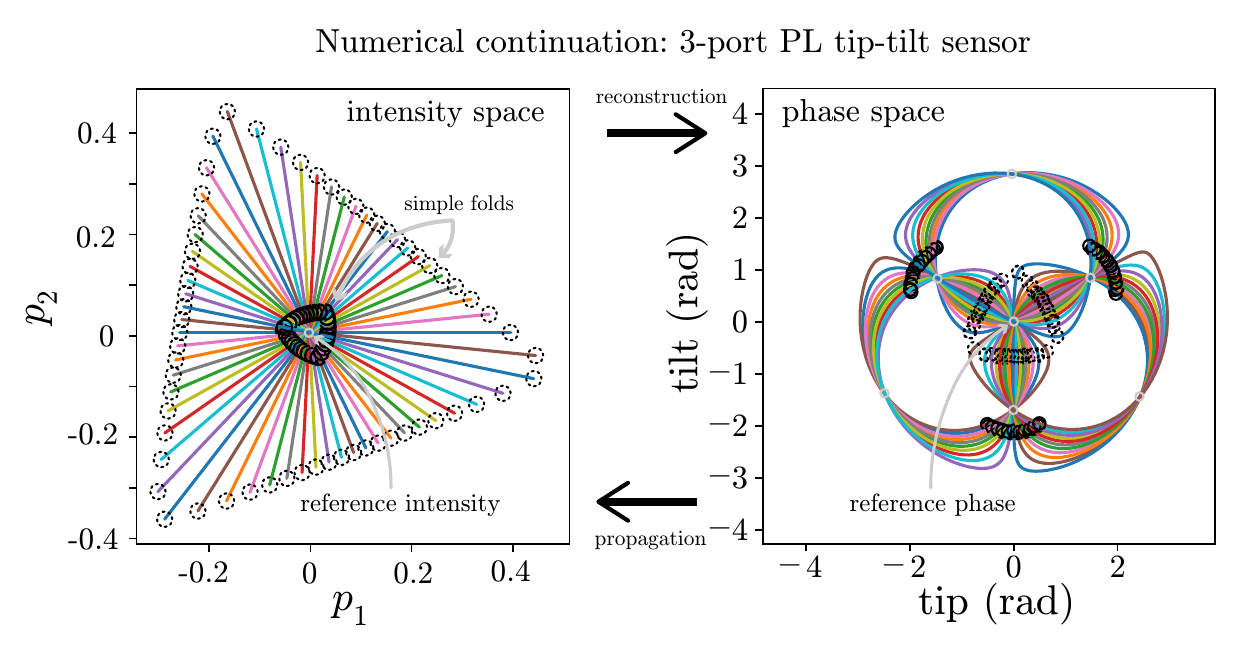}
    \caption{Similar to Figure \ref{fig:nonlinqc} but for the 3-port lantern tip-tilt sensor. Dashed and solid black circles indicate corresponding sets of simple folds, as they appear in intensity and phase space. Line segments in intensity space map to closed curves, and the reference intensity (gray circle in left panel) is mapped to 7 distinct phases (gray circles in right panel), each of which are marked by curve crossings. The presence of phase degeneracies indicate that the function taking intensity to phase is multi-valued; branches of this function are partitioned by simple folds. }
    \label{fig:nonlin}
\end{figure}

\section{Discussion } \label{sec:disc}
\subsection{Feasibility}
The primary barrier to a real-time demonstration of the methods presented in this work is in calibration. For one, the calibration for some methods may take a prohibitively long time. Taylor expansion methods in particular show poor scaling with the number of sensed modes: an $M$-mode sensor calibrated to order $p$ will require $\lesssim M^p$ probes on the deformable mirror. As such the realistic upper-bound for calibration order with $\sim$10-mode sensors is probably 3 or 5 (as mentioned earlier, it is not efficient to truncate the power series at even orders), and the calibration process will take tens of thousands of probes on the deformable mirror. The scaling between the number of required calibration probes and the number of sensed modes is less clear for phase retrieval strategies based on RBF interpolation, since we only have two datapoints for two specific sensors: around 100 control points for a 2-mode quad-cell sensor and 1000 for a 5-mode PL sensor. Future modelling of higher-mode sensors would help determine the exact scaling. For now, under the assumption of a power law scaling, the number of calibration points required for a 10 mode sensor is around 5,000 points, similar to the Taylor expansion models and some neural network schemes \cite{Norris:20}. In any case, the longer calibration times for nonlinear phase retrieval methods likely necessitates stabilization of the telescope and instrument, since any uncorrected and time-varying aberrations will add inconsistency to the nonlinear model; alternatively, we may account for noise using smoothing interpolators or Gaussian process models.
%Lastly, we note that it may be possible to calibrate RBF interpolation models using AO telemetry data.

\subsection{Composite sensing strategies}
The phase retrieval methods presented in this paper may be combined with each other, as well existing methods, to produce more robust and performant sensing schemes. For instance, we might construct a dual-stage sensing scheme which uses RBF-I when the phase error is anticipated to be large and a linear or cubic Taylor method when the phase error is small, which would have an effectively nonlinear dynamic range while maintaining fast computation times and a high degree of accuracy. Such a composite method is similar to the combined neural network + linear methods being tested on the SCExAO testbed at Subaru telescope \cite{kyohoon}. 
\\\\
For another example, in \S\ref{sec:num} we found that the performance of the RBF-I method may suffer if the calibration dataset happens to contain phase aberrations from different branches of the WFS response. Because it may be difficult to exclude problematic control points {\it a priori}, one workaround is to instead calibrate a backwards RBF model in a very small neighborhood of the reference phase, and use this model to calibrate a higher-order Taylor expansion which directly takes intensity to phase. Lastly, simple nonlinear phase retrieval schemes for quad-cells might be chained together to improve the performance of Shack-Hartmann sensors. Analogously, because the response matrices for pyramid WFSs typically have a banded diagonal structure, few-mode nonlinear sensing strategies might also be composited together for the pyramid as well.

\subsection{Forwards or backwards?}
The relative performance between forward methods like RBF-LS and backward methods like RBF-I will depend on the properties of the nonlinear response map $\mathcal{T}$. For instance, RBF-I outperforms the RBF-LS method for the quad-cell sensor, but is outperformed in the case of the PL sensor. This discrepancy is likely due to a difference in where degenerate aberrations occur in the phase space for each sensor. In the case of the quad-cell, we found via numerical continuation (Figure \ref{fig:nonlinqc}) that degeneracies begin to occur when the total RMS WFE exceeds around 2.5 radians RMS. The phase aberrations randomly sampled for calibration all fall within 2 radians RMS, so the RBF-I model is well-behaved. In contrast, the calibration probes for the 5-mode PL sensor extend out to 1 rad RMS total wavefront error, while numerical continuation (see \S\ref{ap:B} and Figure \ref{fig:numcont_5mode}) shows that degeneracies may occur within 1 rad RMS. If the calibration dataset includes aberrations from different branches of the WFS response, then the inverse model will be inconsistent. Even so, it may be beneficial to include degenerate aberrations during the calibration process, because in sampling a larger area of phase space we may gain dynamic range in exchange for accuracy. Still, the optimal strategy is to map each branch of the WFS response separately.

\subsection{Prospects for dispersion}\label{ssec:dispdisc}
\begin{figure*}
    \centering
    \includegraphics[width=\textwidth]{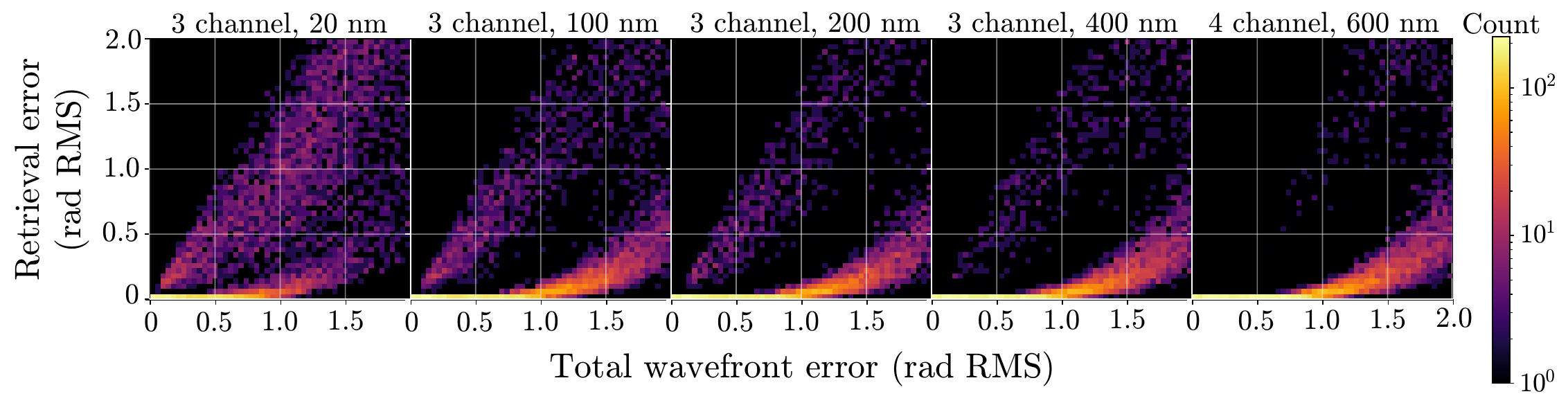}
    \caption{Phase retrieval error for various polychromatic PL sensors, each with different spectral bandwidths. The modal space for each sensor is restricted to the first 7 non-piston Zernike modes. The presence of data points that fall in the vicinity of the line $y=x$ for the 3-channel sensors indicate that they cannot sense a subset of the modal space; this subset becomes smaller when the spectral bandwidth is increased, and mostly disappears for the four-channel sensor.   }
    \label{fig:disp}
\end{figure*}

We can extend monochromatic phase retrieval to polychromatic sensors in the simplest way. Suppose we have wavefront aberrations that are achromatic in optical path difference ($\lambda^{-1}$ chromaticity in phase), and a polychromatic sensor which observes at $L$ distinct wavelengths over the sensing band. We treat this sensor as a function $\mathbb{R}^M\rightarrow \mathbb{R}^{NL}$, and attempt to apply the same techniques developed for monochromatic sensing to this higher-dimensional function. As an example, Figure \ref{fig:disp} shows retrieval error plots for polychromatic sensors using different numbers of wavelength channels and spectral bandwidths; each configuration is composed of several monochromatic 5-mode PL sensors, and uses the RBF-LS reconstruction method. We restrict the modal phase space of these sensors to the first 7 non-piston Zernike modes. All 3-channel sensors observe at a central wavelength of 1.55 $\mu$m while the 4-channel sensor augments the 3-channel, 400 nm bandwidth sensor with an additional channel at 1.15 $\mu$m. We note that at the smallest bandwidths, the residual does not go to zero faster than linear, even for small amounts of WFE, signalling that the sensor is unable to distinguish between certain phase modes. However, as the bandwidth and number of channels increases, so too does correction quality, with the 4-channel configuration showing good performance out to 1 rad RMS. Thus, we have at least shown in this rudimentary test that chromaticity can improve both dynamic range and spatial resolution. We leave more detailed analysis, experimental verification, and the sensing of chromatic aberrations to a followup work.
\\\\
Lastly, we mention one caveat to polychromatic sensing. In the simple construction above we will run into calibration issues when the guide star spectrum changes, since the response in each spectral channel depends on the intensity at each wavelength. If the spectrum changes only slightly, we can correct for the spectral offset in a similar way to how sensitivity loss of the pyramid WFS in the presence of phase offsets is corrected (``optical gain compensation'', e.g. \citenum{chamb}). However, there is no guarantee that the target spectrum will change only slightly. 

\section{Conclusion}
In \S\ref{sec:model}-\S\ref{sec:inv}, we consider two methods for empirically modelling the nonlinear response of few-mode wavefront senors: Taylor expansion and radial basis function interpolation. In conjunction with these models, we also discuss how such models can be solved. In particular we show that higher-order Taylor expansions can be solved via successive approximations, and that forward models of the WFS built with radial basis functions can be solved with standard least-squares techniques as well as numerical continuation. Furthermore, we show that backwards models, which map intensity directly to phase, can be built from radial basis functions as well. \S\ref{sec:num} presents numerical demonstrations of the aforementioned modelling techniques and solving strategies for both a tip-tilt quad-cell sensor as well as a PL WFS which is sensitive to the first 5 non-piston Zernike modes, with additional discussion on how the method of numerical continuation may be used to better understand wavefront sensor nonlinearity and phase aberration degeneracy. Finally, in \S\ref{sec:disc} we discuss future outlooks for both nonlinear and dispersed wavefront sensing.

\section{Backmatter}

\begin{backmatter}
\bmsection{Acknowledgment}
The authors would like to thank the anonymous referee, who provided many helpful comments which improved the manuscript.

\bmsection{Funding}
Content in the funding section will be generated entirely from details submitted to Prism. Authors may add placeholder text in the manuscript to assess length, but any text added to this section in the manuscript will be replaced during production and will display official funder names along with any grant numbers provided. If additional details about a funder are required, they may be added to the Acknowledgments, even if this duplicates information in the funding section. See the example below in Acknowledgements. For preprint submissions, please include funder names and grant numbers in the manuscript.

\bmsection{Disclosures}
The authors declare no conflicts of interest.

\bmsection{Data Availability Statement}
Data underlying the results presented in this paper are not publicly available at this time but may be obtained from the authors upon reasonable request.
\end{backmatter}

\appendix

\section{Pseudo-arclength continuation}\label{ap:pseudo}
Define a potentially vector-valued nonlinear function $\mathcal{F}$ which takes in $M$ algebraic variables $\bm{a}$ and $N$ parameters $\bm{p}$; technically, there is no difference between the two and that some of the variables in $\bm{a}$ may at any time be taken as parameters, or vice versa. The goal is find the set of solutions $\bm{a}$ which satisfy 
\begin{equation}
    \mathcal{F}\left[\bm{a},\bm{p}\right] = 0 
\end{equation}
as $\bm{p}$ moves along some predetermined curve in the parameter space, beginning at some known start point $\bm{a}_0,\bm{p}_0$ for which the above is satisfied. For simplicity, we will consider only the one-parameter case, since any curve can be parameterized in terms of arclength. Therefore, the reduced problem at hand is to numerically continue the solutions $\bm{a}$ which solve
\begin{equation}
    \mathcal{F}\left[\bm{a},p\right] = 0 
\end{equation}
as $p$ varies through some interval $[p_0,p_f]$. We denote the (Fréchet)
derivatives as
\begin{equation}
\begin{split}
    \mathcal{F}_{\bm{a}} = \dfrac{\partial \mathcal{F}}{\partial \bm{a}}  \\
    \mathcal{F}_{p} = \dfrac{\partial \mathcal{F}}{\partial p}
\end{split}
\end{equation}
For a vector-valued $\mathcal{F}$, the first line is the Jacobian with respect to the variables $\bm{a}$. In the simplest numerical continuation schemes, we move $p$ through the interval and repeatedly use a nonlinear solver (e.g. Newton's method) to obtain the solutions $\bm{a}$ as a function of $p$. Each value of $\bm{a}$ is taken as an initial guess for the next value of $p$. However, this process will fail if $\mathcal{F}_{\bm{a}}$ is singular anywhere along the the path, because the nonlinear solver will not converge. Therefore, we will parameterize all arguments of $\mathcal{F}$ in terms of a path length $s$: 
\begin{equation}
    \mathcal{F}\left[\bm{a}(s),p(s)\right] = 0.
\end{equation}
By differentiating the above with respect to $s$ we obtain
\begin{equation}
    \mathcal{F}_{\bm a} \dot{\bm{a}} + \mathcal{F}_{p} \dot{p} = 0
\end{equation}
where dots denote derivative with respect to $s$. We can additionally apply a normalization condition
\begin{equation}
    ||\dot{\bm{a}}||^2 + ||\dot{p}||^2 = 1.
\end{equation}
The combination of the above two formulas determines the tangent vector as 
\begin{equation}
\begin{split}
    \dot{\bm{a}} &= - \mathcal{F}_{\bm{a}}^{-1}\mathcal{F}_p \dot{p} \\
    \dot{p} &= \pm \dfrac{1}{\sqrt{1 + ||\mathcal{F}_{\bm{a}}^{-1}\mathcal{F}_p||^2}}
\end{split}
\end{equation}
which we can evaluate at some starting $\bm{a}_0$ and $p_0$ to obtain the starting tangent vector $[\dot{\bm{a}}_0^T,\dot{p}_0]^T$.
Finally, we augment our nonlinear system as 
\begin{equation}
    \begin{bmatrix}
        \mathcal{F}\left[\bm{a}(s),p(s)\right] \\
        \dot{\bm{a}}_0 \cdot (\bm{a}-\bm{a}_0) + \dot{p}_0(p-p_0)
    \end{bmatrix}
    =
    \begin{bmatrix}
        0 \\
        s-s_0
    \end{bmatrix}
\end{equation}
where the additional scalar equation comes from the normalization condition. The system is solved with initial guess
\begin{equation}
\begin{split}
    \Tilde{\bm{a}} &= \bm{a}_0 + (s-s_0) \dot{\bm{a}}_0  \\
    \Tilde{p} &= p_0 + (s-s_0) \dot{p}_0
\end{split}
\end{equation}
to obtain a solution $\bm{a}$ located a distance $ds=s-s_0$ down the path. This process is then iterated to generate the entire solution curve.
\\\\
Pseudo-arclength continuation allows us to numerically continue through points where $\mathcal{F}_{\bm a}$ is singular so long as we also have $\dot{p}=0$; these points are called simple folds or limit points and correspond to when the path in parameter space folds back on itself. If we do not have $\dot{p}=0$, the solutions may bifurcate along multiple paths, and leads into {\it bifurcation theory}.
\begin{figure}[ht]
    \centering
    \includegraphics[width=\textwidth]{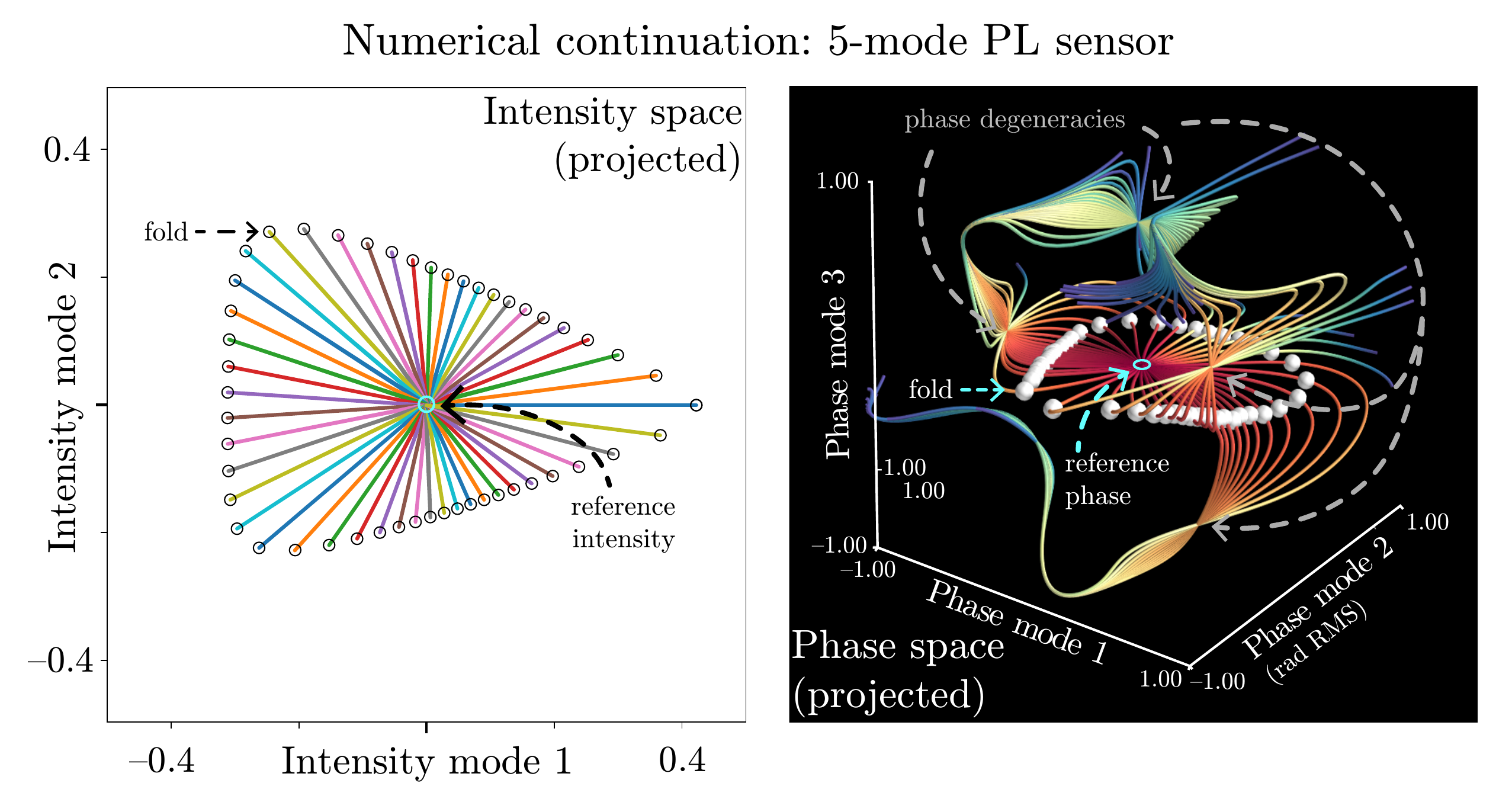}
    \caption{Numerical continuation plot for the 5-mode PL sensor. Intensity space and phase space are represented in the bases corresponding to the left and right singular vectors of the WFS Jacobian; the left and right panels show projections of intensity and phase space. Intensity space is probed along the plane corresponding to the first two modes. In the vicinity of the reference phase, corresponding phase space solution curves fall onto the plane corresponding to the first two phase space modes. As the RMS WFE increases, curves deviate from the plane, indicating nonlinearity. Phase curves are colored by arclength. We note that the reference phase is degenerate to at least 4 other phase aberrations. Simple folds are marked on the left with black circles and on the right by white dots.}
    \label{fig:numcont_5mode}
\end{figure}
\section{Nonlinear characterization by numerical continuation in >3 dimensions}\label{ap:B}
Numerical continuation plots for 2- and 3-mode sensors are straightforward because we can view the entirety of phase space without projection. However, beyond 3 dimensions, locating branches, folds, and degenerate phase aberrations becomes challenging. One way forward is to work in both the left and right singular vectors (the control modes) of the Jacobian. In this basis, the WFS applies a scaling transformation in the neighborhood of the reference phase and intensity. Thus, if we probe intensity space along a plane, solution curves in phase space will also be embedded in a plane, at least near the reference; in other words, with this this approach numerically-continued curves will locally behave like those for 2-mode sensors. As the RMS WFE increases, the phase space surface containing the solutions will bend out of the plane, indicating nonlinearity. As an example, we apply this method to the 5-mode PL sensor from \S\ref{sec:num}, probing intensity space in the plane of the first two intensity modes. These paths are shown in Figure \ref{fig:numcont_5mode} (left); note that some paths were truncated to save on computation time. As before, simple folds in intensity space are marked by open black circles. The WFS relates each path in this 2D intensity space to a solution curve in the 5D phase space. We plot these curves, as projected onto the subspace of the first 3 phase modes, in Figure \ref{fig:numcont_5mode} (right); curves are colored by arclength and simple folds are marked by white dots. Near the origin (corresponding to the reference phase), all curves lie on the plane corresponding to the first two phase modes, because in our chosen basis the WFS acts locally like a scaling transformation. Moving away from the origin, the surface containing the phase solutions bends away from this plane. These curves intersect again at four other locations in phase space, indicating aberrations that yield the same intensity response as the reference. Finally, we note that simple folds begin to appear around 0.5 rad RMS from the reference phase, consistent with the maximum dynamic range of this 5-mode sensor, as shown in Figure \ref{fig:numcomp} (RBF-LS method). Simple folds divide the 2D solution manifold into at least 5 distinct branches, though we might find more if we extend the numerical continuation.

\bibliography{refs}

%%%%%%%%%% If preparing manually:
% \begin{thebibliography}{1}
% \newcommand{\enquote}[1]{``#1''}

% \bibitem{Zhang:14}
% Y.~Zhang, S.~Qiao, L.~Sun, Q.~W. Shi, W.~Huang, L.~Li, and Z.~Yang,
%   \enquote{Photoinduced active terahertz metamaterials with nanostructured
%   vanadium dioxide film deposited by sol-gel method,}
%   {\protect\JournalTitle{Optics Express}} \textbf{22}, 11070--11078 (2014).

% \bibitem{Optica}
% {Optica}, \enquote{{Optica Publishing Group},}
%   \url{http://www.opg.optica.org}.

% \bibitem{FORSTER2007}
% P.~Forster, V.~Ramaswamy, P.~Artaxo, T.~Bernsten, R.~Betts, D.~Fahey,
%   J.~Haywood, J.~Lean, D.~Lowe, G.~Myhre, J.~Nganga, R.~Prinn, G.~Raga,
%   M.~Schulz, and R.~V. Dorland, \enquote{Changes in atmospheric consituents and
%   in radiative forcing,} in \enquote{Climate Change 2007: The Physical Science
%   Basis. Contribution of Working Group 1 to the Fourth assesment report of
%   Intergovernmental Panel on Climate Change,}  S.~Solomon, D.~Qin, M.~Manning,
%   Z.~Chen, M.~Marquis, K.~B. Averyt, M.~Tignor, and H.~L. Miler, eds.
%   (Cambridge University Press, 2007).

% \end{thebibliography}

\end{document}